\newif\ifarxiv
\arxivtrue 
\ifarxiv
\documentclass[preprints,article,accept,moreauthors,pdftex]{Definitions/mdpi_arxiv}
\else 
\documentclass[symmetry,article,accept,moreauthors,pdftex]{Definitions/mdpi}
\fi

\firstpage{1} 
\pubvolume{13}
\issuenum{11}
\articlenumber{2157}
\pubyear{2021}
\copyrightyear{2021}
\externaleditor{Academic Editor: Toshio Tagawa} 
\datereceived{6 September 2021} 
\dateaccepted{29 October 2021} 
\datepublished{11 November 2021} 
\hreflink{https://doi.org/10.3390/\linebreak sym13112157}

\ifarxiv
\pdfoutput=1
\fi

\usepackage[normalem]{ulem}

\newcommand{\be}{\begin{equation}}
	\newcommand{\ee}{\end{equation}}
\newcommand{\bea}{\begin{eqnarray}}
	\newcommand{\eea}{\end{eqnarray}}
\newcommand{\bel}{\begin{align}}
	\newcommand{\eel}{\end{align}}

\def\p{\partial}

\def\P{{\rm P}}

\def\eps{\epsilon}

\definecolor{cyan}{rgb}{0,0.9,0.9}
\definecolor{orange}{rgb}{0.9,0.5,0}
\definecolor{magenta}{rgb}{1,0,1}
\definecolor{purple}{rgb}{0.8,0.4,0.8}
\definecolor{gray}{rgb}{0.8242,0.8242,0.8242}
\definecolor{light-gray}{gray}{0.95}

\Title{Machine Learning for Conservative-to-Primitive in Relativistic~Hydrodynamics}

\TitleCitation{Machine Learning for Conservative-to-\linebreak Primitive in Relativistic Hydrodynamics}

\Author{{Tobias Dieselhorst}
	$^{1,}$*\orcidA{}, William Cook $^1$\orcidB{}, Sebastiano Bernuzzi $^1$\orcidC{} and David Radice $^{2,3,4}$\orcidD{}}

\AuthorNames{Tobias Dieselhorst, William Cook, Sebastiano Bernuzzi, and David Radice}
\AuthorCitation{Dieselhorst, T.; Cook, W.; Bernuzzi, S.; Radice, D.}

\address{%
	$^{1}$ \quad Theoretisch-Physikalisches Institut, Friedrich-Schiller-Universit{\"a}t Jena, 07743 Jena, Germany; william.cook@uni-jena.de~(W.C.); sebastiano.bernuzzi@uni-jena.de~(S.B.)\\
	$^{2}$ \quad Institute for Gravitation \& the Cosmos, The Pennsylvania State University, University Park, PA 16802, USA; dur566@psu.edu \\
	$^{3}$ \quad Department of Physics, The Pennsylvania State University, University Park, PA 16802, USA\\
	$^{4}$ \quad Department of Astronomy \& Astrophysics, The Pennsylvania State University, \mbox{University Park, PA~16802, USA}}

\corres{Correspondence: tobias.dieselhorst@uni-jena.de}

\abstract{The numerical solution of relativistic hydrodynamics equations in conservative form requires root-finding algorithms that invert the conservative-to-primitive variables map. These algorithms employ the equation of state of the fluid and can be computationally demanding for applications involving sophisticated microphysics models, such as those required to calculate accurate gravitational wave signals in numerical relativity simulations of binary neutron stars. This work explores the use of machine learning methods to speed up the recovery of primitives in relativistic hydrodynamics. Artificial neural networks are trained to replace either the interpolations of a tabulated equation of state or directly the conservative-to-primitive map. The application of these neural networks to simple benchmark problems shows that both approaches improve over traditional root finders with tabular equation-of-state and multi-dimensional interpolations. In particular, the neural networks for the conservative-to-primitive map accelerate the variable recovery by more than an order of magnitude over standard methods while maintaining accuracy. Neural networks are thus an interesting option to improve the speed and robustness of relativistic hydrodynamics algorithms.}


\begin{document}


\section{Introduction}
\label{sec:intro}

Relativistic hydrodynamics is often written as a set of conservation
laws (with source terms)
\be \label{eq:hyperbolic}
\frac{\partial\mathbf{u}}{\partial t} + \frac{\partial \mathbf{F}^i(\mathbf{u})}{\partial x^i} = \mathbf{s}\,, 
\ee
where the state vector $\mathbf{u}=\left(D,S_i,\tau\right)$ is composed of the
conserved variables: rest-mass density, momentum density and the
energy density relative to $D$, $\tau:=E-D$, all measured in the
laboratory frame~\cite{Marti.2003} {($i=1,2,3$)}~(we
focus here on Cartesian coordinates and special relativity,
although the main results of the paper are directly applicable to
arbitrary coordinate system and general relativistic hydrodynamics).
Conserved variables are related to primitive variables,
$\mathbf{w} = \left(\rho, v^i, \epsilon,p\right)$,
defined in the local rest frame of the fluid through (in units of
light speed $c=1$)
\be\label{eq:p2c}
D = \rho W,\ \ \
S_i = \rho h W^2 v_i, \ \ \ 
\tau = \rho h W^2 - p - D,
\ee
where $\rho$ is the rest-mass density, $v^i$ the fluid's 3-velocity, 
$\epsilon$ the specific internal energy (the total energy is
$e=\rho(1+\epsilon)$), {$p$ is the pressure, }
$h=1+\epsilon+p/\rho$ is the specific enthalpy and
$W=(1-v^2)^{-1/2}$ is the Lorentz factor between the two reference
frames. {$\mathbf{F}^i = (Dv^i, S^1v^i+p\delta^{1i}, S^2v^i+p\delta^{2i}, S^3v^i+p\delta^{3i}, S^i-Dv^i)$ are the flux vectors.}
The conservative-to-primitive
transformation $\mathbf{w}(\mathbf{u})$ inverting \eqref{eq:p2c} 
cannot be written in closed form.
The system \eqref{eq:hyperbolic} is closed with an equation of state
(EOS). In its simplest form, the EOS is the thermodynamical relation
connecting the pressure to the fluid's rest-mass density and internal energy,
$p=\bar{p}(\rho,\epsilon)$. If more particle species are considered,
then the EOS also includes additional thermodynamical potentials. For
example, the Helmholtz and Lattimer--Swesty EOS used in various
relativistic astrophysics simulations~\cite{Timmes.2020,Lattimer:1991nc}
return the pressure as function of the rest-mass density (or baryon
number density), the temperature and the electron fraction, i.e., $p=\bar{p}(\rho,T,Y_e)$.
In this case, Equation~\eqref{eq:hyperbolic} can also contain
additional continuity equations for the rest-mass or 
number densities of the different species (e.g., for the variable $DY_e$).
For a causal EOS, the system \eqref{eq:hyperbolic} is hyperbolic
\cite{Anile.1989}. The conservative form of the equations allows the
application of robust mesh-based numerical methods that correctly capture weak
solutions, e.g., shocks~\cite{Toro.1999,Leveque.2002}.
In an astrophysical context, these high-resolution shock-capturing (HRSC) methods are
routinely employed for the simulation of relativistic jets
\cite{Marti.2003,Blandford:2018iot}, supernova explosions~\cite{Janka:2012wk}
and binary neutron star mergers in general relativity~\cite{Font:2007zz,Radice:2020ddv,Bernuzzi:2020tgt}.

A central algorithmic step in hydrodynamics codes is the computation
of the \linebreak conservative-to-primitive (C2P) inversion. This calculation is
performed by applying root finders at each mesh point and for each
r.h.s. evaluation during the time evolution and it involves an evaluation of the EOS.
In multi-dimensional simulations including microphysics,
this procedure can become cumbersome and can lead to numerical errors and failures, 
see e.g., discussions in~\cite{Noble:2005gf,Siegel:2017sav,Kastaun:2020uxr}.
For example, general relativistic simulation of neutron star mergers
with microphysics typically employ 
a C2P procedure composed of two nested root finders (Newton--Raphson
algorithms) during which 3D interpolations of EOS tables are performed
to find the searched-for variable (e.g., the pressure
$\bar{p}(\rho,T,Y_e)$), e.g.,~\cite{Sekiguchi:2015dma,Foucart:2016rxm,Radice:2018pdn}.
This procedure is called at each grid point for each time subcycle,
resulting in more than $10^{9}$ calls per millisecond of
evolution in simulations that currently span up to hundreds of
milliseconds~\cite{Radice:2018pdn,Bernuzzi:2020txg,Nedora:2020pak}.
The computational cost of the C2P amounts up to 
${\sim}18\%$ of the total cost for computing the r.h.s. of general
relativistic equations in hyperbolic form. The C2P also
impacts parallel computations and produces load imbalances since
different mesh points (ranks) require different workloads.
Moreover, the C2P computation is 
memory intensive as about $300$~MB of EOS data must be loaded for these
operations.  
A key goal of numerical relativity is to use such simulations to generate
highly accurate gravitational wave signals. These are used to detect binary
neutron star mergers in gravitational wave detectors such as the Laser Interferometer
Gravitational Wave Observatory (LIGO) and then to perform parameter estimation to
extract the parameters of the neutron stars from the wave signal. Improving the 
performance of simulations  such as those detailed above will allow numerical relativity codes
to generate more accurate gravitational waveforms more efficiently.

This paper explores a new approach to the conservative-to-primitive
inversion based on supervised learning. Specifically, the use of
artificial neural networks is investigated in order to produce fast
algorithms that can substitute either a multi-dimensional EOS
interpolation or the entire transformation of variables without affecting
the accuracy of the solution. 
Section~\ref{sec:method} summarizes the construction of simple neural
networks for the EOS and the C2P adopting an analytical $\Gamma$-law
EOS as a benchmark,
\be\label{eq:eos}
\bar{p}(\rho,\epsilon) = (\Gamma-1)\rho\epsilon\,.
\ee

Section~\ref{sec:results} discusses the performances of the neural
networks in terms of speed and accuracy by comparing them against the results of
the C2P performed with a standard inversion algorithm and using both the
analytical and a 3D tabulated representation of the EOS. The
performances of the neural networks are also evaluated on two simple
test problems for 1D (special) relativistic hydrodynamics: a relativistic shock
tube and a smooth solution.
In Section~\ref{sec:conc} we conclude that neural network representations 
of the C2P are an interesting option to improve the speed and robustness
of relativistic hydrodynamics. 
Appendix~\ref{app:c2p:root} reviews a standard method to compute the
conservative-to-primitive inversion using root finders.

\section{Method}\label{sec:method}

The basis of any neural network (NN) is the neuron that, just like its
biological counterpart, receives input values, builds a weighted sum
over these input values, adds a bias and then generates an output by
applying a (nonlinear) activation function, $\sigma$, to the sum; see
e.g.,~\cite{Goodfellow.2016,Mehta.2019,Schmidhuber.2015} for an introduction on NNs.
A feedforward NN contains multiple layers of such neurons. The first and last 
layers are called the input and output layer respectively, while the
intermediate layers are called hidden layers.
In case of a fully connected NN, the output of each layer 
becomes the input for all neurons in the next layer. The output of the
output layer is handed back as the output of the
neural network. 

All NNs used for this work are fully connected feedforward neural
networks with two hidden layers.
The NN parameters are the weights and biases and are collectively indicated as $\theta$.
They are determined in the training step by minimizing the difference
between the NN's output, $\hat{y}_i(\theta)$, and the exact values
($y_i$), called ``labels'' of a training dataset.
The minimization is performed iteratively by (i) computing a loss
function $E$, (ii) taking the gradient of $E$ with respect to all parameters using the chain rule (backpropagation step), (iii) applying a gradient
descent algorithm to adjust all parameters in order to minimize $E$. An epoch is completed when
steps (i--iii) are performed for all samples of the training dataset.
Here, the training data are split into random mini-batches and the
gradients of the parameters $\theta$ are collected for all
samples of a mini-batch. The gradient descent algorithm is then
applied after each mini-batch. New mini-batches are created randomly
after each epoch. 
The specific loss function minimized in the training process is 
\begin{align} \label{eq:MSE}
	E(\theta) = \frac{1}{n}\sum\limits_{i=1}^{n} (\hat{y}_i(\theta)-y_i)^2\,,
\end{align}
and the Adam optimizer is used for the gradient descent~\cite{Kingma.22.12.2014}. 

For the construction of each NN we experimented
with different hyperparameters (e.g.,\ number of layers, layer sizes,
activation functions). In particular, we considered NNs with two and three
hidden layers and layer sizes from $100$ to $1200$.
The considered activation functions are $\tanh(z)$ and
\begin{align}
	\mathrm{ReLU}(z) &= \max(0,z) \label{eq:ReLU}\\
	\mathrm{Sigmoid}(z) &= \frac{1}{1+e^{-z}}. \label{eq:Sigmoid}
\end{align}

The NNs proposed and described below hit a balance between simplicity, accuracy and size.
The measured improvement in accuracy of larger NNs was only marginal in contrast to a longer evaluation time. Similarly, the use of three hidden layers resulted in minor improvements over two layers for the
same size while increasing training and evaluation time
significantly. The best results were achieved with the
$\mathrm{Sigmoid}$ activation function (Equation~\eqref{eq:Sigmoid}) as the
activation function of the hidden layers and \mbox{$\mathrm{ReLU}$~(\ref{eq:ReLU})} as the nonlinearity applied on the output.
In order to evaluate the the NNs' performance, a test dataset is used
which is held back from the training. Note that overfitting the
training data did not occur in our tests, as all our networks have a lower \mbox{loss  (\ref{eq:MSE})} on an independent (smaller) testing dataset than on the training dataset. Thus, e.g.,\ dropout methods do not improve the training.
All algorithms described in this paper are implemented in {\tt Python} and NNs
are implemented with {\tt PyTorch}~\cite{Paszke.2019}. All implementations are vectorized and make use of {\tt NumPY}~\cite{harris2020array} optimizations. We note, however, that the NN implementations are expected to be more optimized than, e.g.,\ the root-finding algorithms. {For implementations of this approach within production codes (e.g.,\ those for general relativistic hydrodynamics) it will be necessary to interface with compiled languages such as {\tt C++}. The evaluation of neural networks, however, only requires the computation of the activation functions and basic linear algebra operations, for which there are several optimized routines available. The training of the NNs is completed beforehand and can be done in any programming language, as only the parameter values (i.e.,\ weights and biases) have to be imported to later applications. Since {\tt PyTorch} as well as other widely used libraries such as {\tt TensorFlow} have {\tt C++} APIs it will also be possible to implement the evaluation of NNs in these codes, after training them using a {\tt Python} code. Libraries such as {\tt cuDNN} can further be used in codes designed for GPUs.}

Throughout this paper, errors on data series are evaluated using the two norms
\begin{align}
	L_1 &:= \frac{1}{n} \sum_{i=1}^n \left|\hat{y}_i - y_i\right|,
	\label{eq:MAE} \\
	L_\infty &:= \max_{i}|\hat{y}_i - y_i| \label{eq:maxE}\,,
\end{align}
where $\hat{y}_i$ are the outputs of the NN (or any derived quantity)
and $y_i$ are the exact values.

\subsection{NN for EOS}

As a first approach, NNs are built to represent the EOS in
Equation~\eqref{eq:eos}. These NNs output the pressure for given
values of the rest-mass density and specific internal energy. We also
consider an example of NN that additionally outputs the derivatives $\chi:=\p p/\p \rho$ and
$\kappa:=\p p/\p \epsilon$. 
The latter are usually necessary to compute the speed of
sound {(see \linebreak (\ref{eq:soundspeed}))}, which is crucially required for the computation of the
numerical fluxes in HRSC algorithms.
The derivatives of pressure (or other thermodynamical potentials) are usually
provided also by microphysical EOS tables and can therefore be used as labels
during the training~process.  

For later applications in Section~\ref{sec:results}, we fix the adiabatic
exponent in Equation~\eqref{eq:eos} to $\Gamma=5/3$ and create a training
set by randomly sampling the EOS on a uniform distribution over
$\rho\!\in\!(0, 10.1)$ and $\epsilon\!\in\!(0, 2.02)$. 
{This range of primitives is specific to the problems described in Section~\ref{sec:results}, hence the NNs used are specifically trained for these problems. When applying these techniques to other problems, where the primitive ranges may differ considerably, the NN should be trained on appropriately sampled data that matches the range of primitives expected in that problem.}
The training dataset has size of $80{,}000$, and it is split into
random mini-batches of 32 samples each. 
The test dataset has size $10{,}000$. 

Two different NN representations of the EOS are investigated:
\begin{description}[leftmargin=50pt,labelsep=5pt]
	\item[NNEOSA] {Type}
	A NN only outputs the pressure. Its gradient can be computed using the backpropagation algorithm, but the error of the derivatives is not taken into account during the training process. 
	\item[NNEOSB] Type B NN has three output neurons, for the pressure and its derivatives respectively. Therefore, the loss of all three values is minimized simultaneously during training.
\end{description}

\begin{figure}[H]
	\centering
	\includegraphics[width=.35\linewidth]{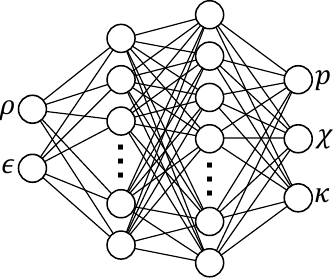}
	\hspace{.1\linewidth}
	\includegraphics[width=.35\linewidth]{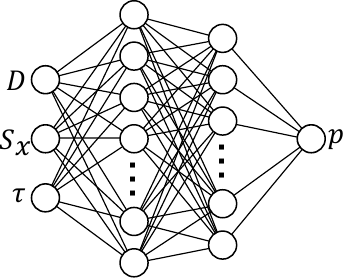}
	\caption{{Schematic diagram of NNEOSB (\textbf{left}) and NNC2P (\textbf{right}). Input layers are on the left of each network and output layers are on the right. NNEOSA is identical to the scheme of NNEOSB but without the output neurons for $\chi$ and $\kappa$.}}
	\label{fig:NNs}
\end{figure}

The NNEOS layouts for each type are summarized in
Table~\ref{tab:NNconfigurations} {and NNEOSB is visualized on the left of Figure \ref{fig:NNs}}. 
{For the training, an initial learning rate of the Adam optimizer~\cite{Kingma.22.12.2014} is set to {$1\cdot10^{-4}$ ($6\cdot10^{-4}$)} for the training of NNEOSA (NNEOSB) and automatically adapted when saturations of the error on the training dataset occurred. The training is completed when the training error does not continue to decrease even after learning rate adaptions} {(see Appendix \ref{app:training} for an example of the training of NNEOSA).}
Comparing to the test dataset {after training}, the difference in the error of the returned pressure $p$ is within the same order of magnitude for both
networks, but it is approximately twice as large for NNEOSA in both the $L_1$ and $L_\infty$ norms. As expected, NNEOSB returns much more accurate values for the derivatives of the pressure as, unlike NNEOSA, it is specifically trained to return these as outputs. As measured in the $L_1$-norm ($L_\infty$-norm), we find that the error in the derivative $\chi$ in NNEOSB is a factor of ${\sim}12$ ($354$) smaller than that of NNEOSA and a factor of $\sim$34 ($124$) smaller for the derivative $\kappa$.

\begin{table}[H]
	\centering
	\caption{Hyperparameters
		and errors for the NNs considered in this work.
		All NNs use two hidden layers (H.L.), both using a Sigmoid activation function, while all output layers use a ReLU activation function.
		The respective test datasets are used to evaluate the accuracy of the representations. Both error norms are shown.}
	\begin{tabular}{ccccccccc}
		\toprule
		\multirow{2}{*}{\textbf{Network}} &  \multicolumn{2}{c}{\textbf{Size of H.L.}} & \multicolumn{2}{c}{\textbf{\boldmath{Error $\Delta p$}}} & \multicolumn{2}{c}{\textbf{\boldmath{Error $\Delta \chi$}}} & \multicolumn{2}{c}{\textbf{\boldmath{Error $\Delta \kappa$}}} \\
		& \textbf{H.L.~1} & \textbf{H.L.~2} & \textbf{\boldmath{$L_1$}} & \textbf{\boldmath{$L_\infty$}} & \textbf{\boldmath{$L_1$}} & \textbf{\boldmath{$L_\infty$}} & \textbf{\boldmath{$L_1$}} & \textbf{\boldmath{$L_\infty$}} \\
		\midrule
		
		NNEOSA & 600 & 300 & $2.13\cdot10^{-4}$ & $6.16\cdot10^{-3}$ & $6.03\cdot10^{-4}$ & $1.81\cdot10^{-1}$ & $2.58\cdot10^{-3}$ & $1.15\cdot10^{-1}$ \\
		NNEOSB & 400 & 600 & $1.03\cdot10^{-4}$ &  $2.52\cdot10^{-3}$ & $4.97\cdot10^{-5}$ & $5.11\cdot10^{-4}$ & $7.66\cdot10^{-5}$ & $9.28\cdot10^{-4}$\\
		NNC2PS & 600 & 200 & $3.84\cdot10^{-4}$ & $8.14\cdot10^{-3}$ & - & - & - & - \\
		NNC2PL & 900 & 300 & $3.62\cdot10^{-4}$ & $9.26\cdot10^{-3}$ & - & - & - & - \\
		\bottomrule
	\end{tabular}
	\label{tab:NNconfigurations}
\end{table}
	
\subsection{NN for C2P}

As a second approach, NNs are built to represent the entire C2P
transformation. These NNs output the pressure $p$ from the three conserved quantities in $\mathbf{u}$.
All the other primitive quantities are then computed analytically from
the conserved variables and the pressure, see Appendix~\ref{app:c2p:root}.

The EOS parameter is again $\Gamma=5/3$. The training dataset is
constructed by uniformly sampling the primitive variables over the
following intervals $\rho \in (0,10.1), \eps\in(0,2.02) , v_x \in
(0,0.721)$, and then calculating the corresponding conservative
variables through Equation~(\ref{eq:p2c}). The training and
test datasets contain $80{,}000$ and
$10{,}000$ samples respectively, and random mini-batches of a size of $32$ were
used for the training process. 

When choosing an NN architecture, a balance between accuracy and
speed must be found. In order to investigate the differences in larger
versus smaller NN, two NN representations of the C2P  differing in the number of
neurons are considered:
\begin{description}[leftmargin=49pt,labelsep=6pt]
	\item[NNC2PS] Type S NN outputs the pressure from the conservative
	variable using a small number of neurons that still guarantee
	pressure errors of order ${\sim}10^{-4}$, as for the NNEOS. 
	\item[NNC2PL] Type L NN has a larger size but it is otherwise
	identical to NNC2PS.
\end{description}

{The networks' architecture is visualized on the right of Figure \ref{fig:NNs}.}
The NNC2P layout for each type is summarized in
Table~\ref{tab:NNconfigurations}. {They are trained with a learning rate of $6\cdot10^{-4}$ and learning rate adaptions until the error on the training dataset is minimized} {(an example of the training of NNC2PS can be found in Appendix \ref{app:training}).}
Again we measure the error in the calculated pressure after the {trained} NN performs the C2P by comparing to a test dataset. Here we find that both NNC2PS and NNC2PL perform very similarly, with an NNC2PL returning an error smaller by a factor of $1.06$ in the $L_1$ norm. We also note that this error is consistent in size with the errors in the pressure produced by NNEOSA and NNEOSB, being the same order of magnitude, with NNEOSB returning errors a factor ${\sim}4$ ($3.5$) smaller than NNC2PS (NNC2PL) in the $L_1$-norm. Since the NN performs the entire C2P, we do not need to perform the intermediate step of calculating the derivatives of the pressure in this case.

\section{Results}\label{sec:results}

In this section we present comparative timing and accuracy tests of
the developed NNs.
Section~\ref{sec:res:c2p} directly considers different C2P algorithms
using NNs and compares them against a standard algorithm for the C2P
used in current numerical relativity simulations of neutron stars
spacetimes~\cite{Radice:2018pdn,Bernuzzi:2020txg,Nedora:2020pak}.
Sections~\ref{sec:res:st} and \ref{sec:res:sw} consider the use of 
different C2P algorithms with NNs in two standard 1D benchmarks for relativistic
hydrodynamics implementations: a relativistic shock tube (problem 1 of~\cite{Marti.2003}) and
the evolution of a smooth density profile. 

In order to properly evaluate the benefits of NN representation of the
EOS and C2P in a realistic and yet controlled benchmark, the EOS in
Equation~\eqref{eq:eos} is stored as a table and evaluated via an eight
point (3D) linear Lagrangian interpolation~\cite{OConnor.2010}. The tables contain a
logarithmic grid of 500 values of the density $\rho$ and the
temperature $T$~(%
{the} specific internal energy is tabulated from $\epsilon =
1/(\Gamma-1)\,N_Ak_B T$, where $N_A$ is the Avogadro number and $k_B$
the Boltzmann constant)
and a linear grid of 500 dummy values of the electron
fraction $Y_e$. The grids are of the same range as the training
dataset of the NNs, i.e.,\ with maximum values 10\% over the maxima of
the shock tube test problem. The derivatives needed for the
root-finders are taken from the linear interpolation.

The following tests are performed using either a standard C2P procedure based on
Newton--Raphson (NR) algorithm and described in Appendix~\ref{app:c2p:root}
(hereafter referred to as the \emph{NR algorithm}) with the analytical, tabular, NNEOSA or NNEOSB 
representation of the EOS or the NNC2PS and the NNC2PL representation
of the C2P procedure. A tolerance level of $10^{-8}$ is employed in
the Newton--Raphson algorithm.

\subsection{Accuracy and Timing of C2P}\label{sec:res:c2p}

The accuracy of the NNs is evaluated on
linear sampling $(\rho,\epsilon)$ on ranges $\rho\in[0.05,10]$ and $\epsilon\in[0.01,2]$ with $n=200$,
inverting the conservatives with
varius C2P algorithms and comparing to the exact values. The considered
C2P algorithms are:
\begin{itemize}
	\item The NR algorithm with the analytical EOS;
	\item The NR algorithm with the EOS in tabulated form;
	\item The NR algorithm with the NNEOSA and NNEOSB representation of the EOS; 
	\item The NNC2PS and NNC2PL algorithms.
\end{itemize}

\begin{figure}[!h]
	\centering
	\includegraphics[width=.82\linewidth]{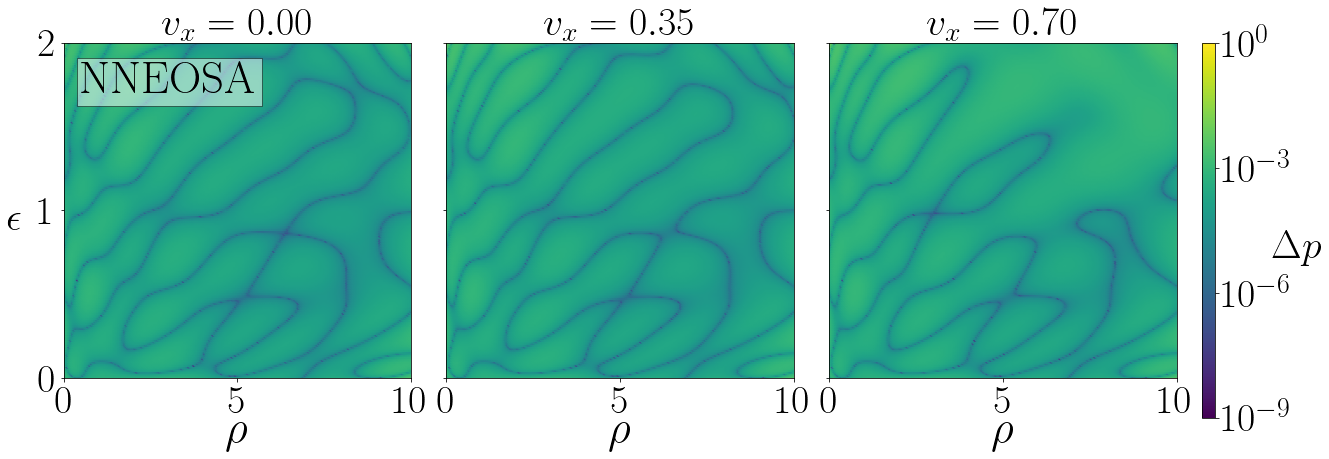}\\
	\includegraphics[width=.82\linewidth]{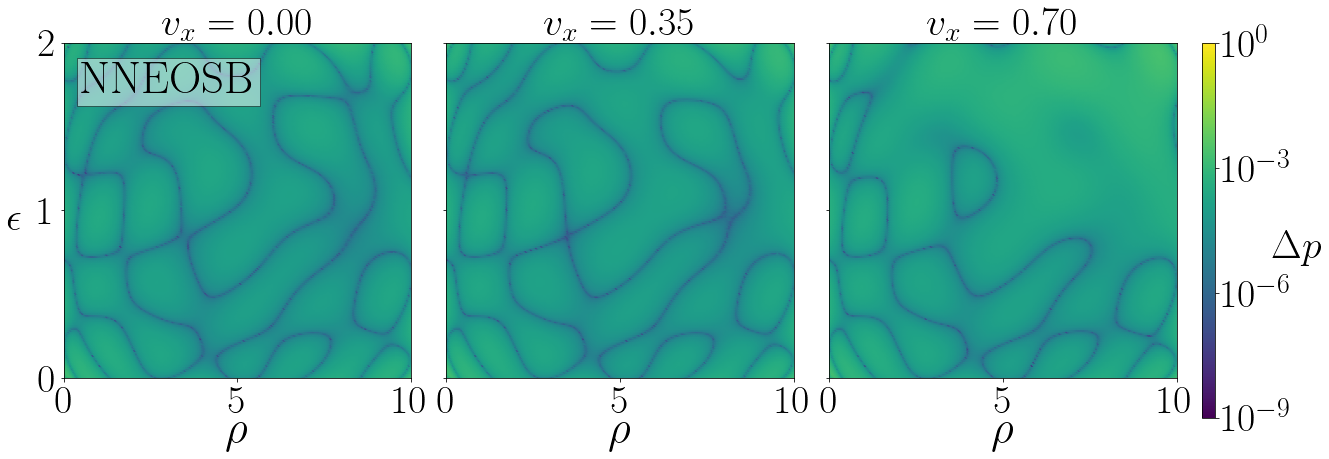}
	\caption{Accuracy of the NNEOS. Discrete per point evaluation of the
		accuracy of the NN EOS representations on a linear grid of $\rho$
		and $\epsilon$ values for three values of the velocity
		$v_x$. The colors show the absolute deviation from
		the exact values (Equation~\eqref{eq:eos}). \textbf{Top} (\textbf{bottom})
		panels refer to NNEOSA (NNEOSB).}
	\label{fig:NNEOS:accuracy}
\end{figure}

\begin{figure}[!h]
	\centering
	\includegraphics[width=.82\linewidth]{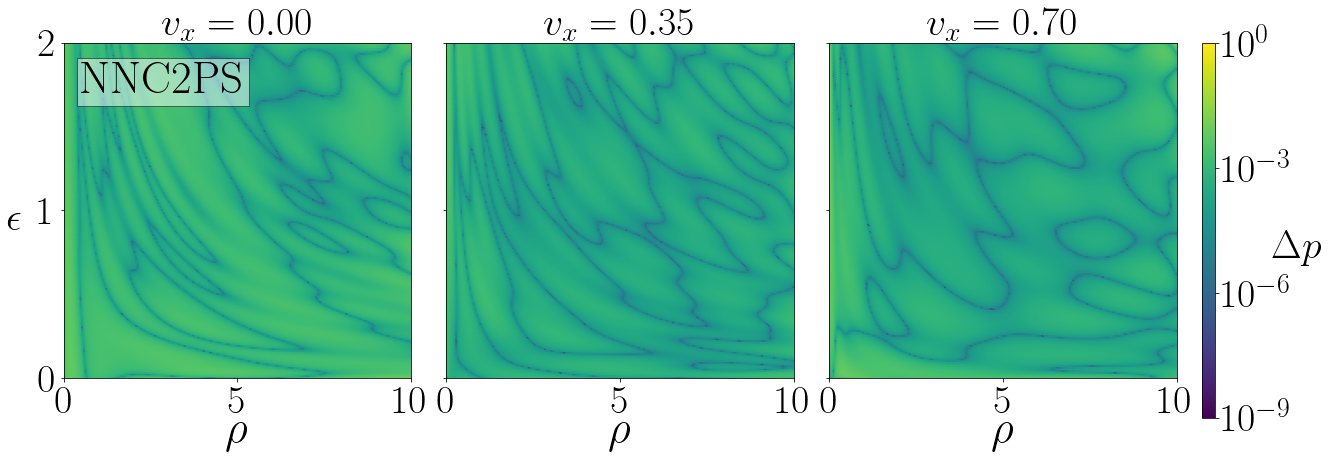}\\
	\includegraphics[width=.82\linewidth]{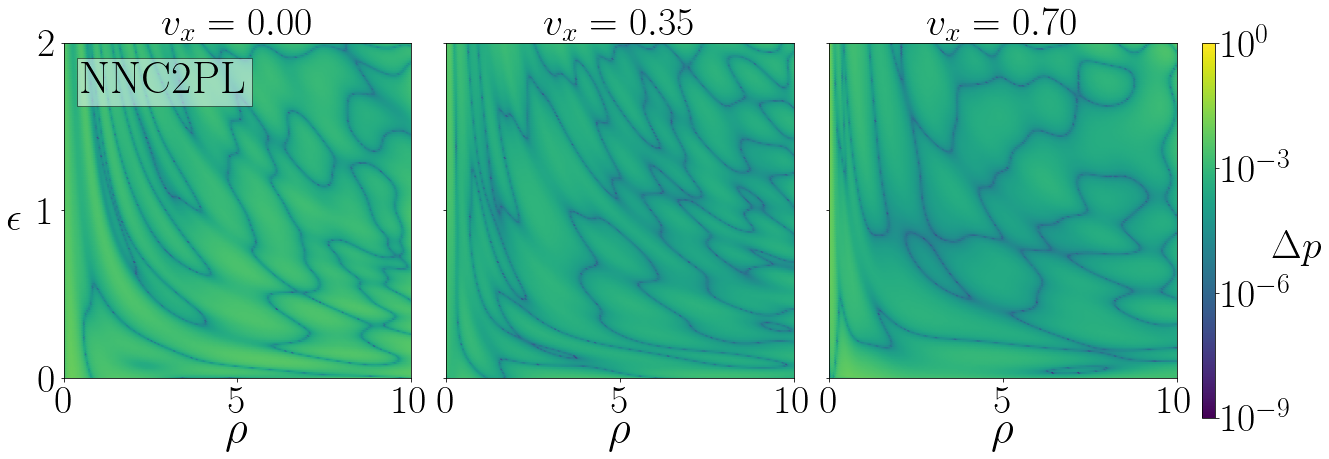}
	\caption{Accuracy of the NNC2P. Discrete per point evaluation of the
		accuracy of the NN C2P representations on a linear grid of $\rho$
		and $\epsilon$ values for three values of the velocity
		$v_x$. The colors show the absolute deviation from
		the exact values (Equation~\eqref{eq:eos}). \textbf{Top} (\textbf{bottom})
		panels refer to NNC2PS (NNC2PL).}
	\label{fig:NNC2P:accuracy}
\end{figure}

Figures~\ref{fig:NNEOS:accuracy} and~\ref{fig:NNC2P:accuracy}
report the absolute errors in the recovery of pressure as a function
of $(\rho,\epsilon)$ in the testing datasets for the NR algorithm
with the NN EOS representation and the NNC2P algorithms respectively. 

The best attainable accuracy is given by the tolerance of the NR algorithm, $10^{-8}$, when taken with the analytical EOS. When combined with the tabulated EOS, we find that the errors grow to an average value of $1.4\cdot 10^{-3}$ for all three velocity values considered in Figures \ref{fig:NNEOS:accuracy} and \ref{fig:NNC2P:accuracy}. The maximum error reaches $1.04\cdot 10^{-2}$ for the highest velocity $v_x = 0.7$, and it is dominated by the contribution from the table interpolation. The accuracy of the NNs proves superior to the tabulated EOS.
Firstly we consider the NNs for evaluation of the EOS coupled with the NR algorithm. For the range of velocities shown NNEOSA has an average error ranging from $2.03\cdot 10^{-4}$ to $3.44\cdot 10^{-4}$, while NNEOSB ranges from $9.80\cdot 10^{-5}$ to $2.34\cdot 10^{-4}$. The latter is an order of magnitude smaller than the error with tabulated EOS. As expected, NNEOSB performs up to a factor of two better than NNEOSA. The maximum error of NNEOSA is comparable to the tabulated EOS; however, the maximum error of NNEOSB is almost an order of magnitude better ($2.29\cdot 10^{-3}$). Again, the size of the errors here demonstrate that the tolerance of the NR algorithm is much lower than the contribution to the error from the NN evaluation of the EOS. 
Secondly we consider the NNs which evaluate the total C2P, NNC2PS and NNC2PL. Here we find the range of average errors for NNC2PS $3.53 \cdot 10^{-4}$--$9.73\cdot 10^{-4}$ and NNC2PL $2.99\cdot 10^{-4}$--$9.25\cdot 10^{-4}$ to be up to an order of magnitude superior to the tabulated EOS and at worst a factor of ${\sim}1.4$ better. The maximum error for either NN is found for the highest velocity in NNC2PL of $9.66\cdot 10^{-3}$, ${\sim}10$\% better than the tabulated EOS. As expected we find that NNC2PL has slightly smaller average errors than NNC2PS, reducing errors between a factor of 1.05 and 1.18 in the three representative velocities considered.  
Finally, we compare the two sets of NNs to each other. As it can be seen from the ranges quoted above, the average errors of all four NNs are of a consistent order of magnitude $\sim 10^{-4}$, with the lowest errors in the EOS, rather than C2P NNs, with NNEOSB performing between a factor of 1.8 and 9.9 better than the C2P NNs in terms of average error.

Timing tests are performed in a similar manner to accuracy tests
but with building grids of states of increasing size from $n=100$ to $n=3200$.
The results are summarized in Table~\ref{tab:C2P:timing}.
As expected, all NN representations perform significantly slower than
the NR algorithm with analytic EOS. Moreover, the cost of all the
algorithms that do not employ the analytical EOS representation
increases proportionally to the grid size.
All the algorithms employing NN representations perform significantly faster than the NR algorithm with tabulated EOS. This is primarily due to the nested iteration of the NR scheme, see Appendix~\ref{app:c2p:root}.
The NR algorithm using the NNEOSB runs up to two times faster than the
same with NNEOSA, despite the fact that NNEOSB is a larger NN.
This is due to the different computations of the derivatives of $p$: while NNEOSB returns the pressure and its derivatives as outputs, NNEOSA 
returns only the pressure, and then the pressure derivatives must be calculated
from backpropagation. This latter procedure increases the run time of the
evaluation. e.g.,\ for $n=800$ gridpoints, generating an output with NNEOSB takes on average $6.07\cdot10^{-3}$~s. While NNEOSA only needs $4.61\cdot10^{-3}$~s to compute $p$, an additional $5.68\cdot10^{-3}$~s are needed to compute the derivatives via backpropagation.
Following expectations, the NNC2PL is slower (by on average a factor of more than 1.5)
than the NNC2PS. While faster NN EOS representations are always
achievable by limiting the network size, this can compromise
accuracy. 
The most significant result is however the comparison of the NNC2P with
the NR algorithm using the interpolated EOS. The NNC2PL completes the
C2P on average ${\sim}13$ times faster than the NR algorithm with
interpolated EOS. NNC2PS is instead ${\sim}22$ times faster.
Thus, these NNs accelerate the C2P by more than an order of magnitude
without significantly compromising the accuracy. We note also that, for all grid sizes, both NNC2P approaches are always faster than both NNEOS approaches, with the faster EOS approach (NNEOSB) slower than the slowest C2P approach (NNC2PL) by at least a factor of 1.7.

\begin{table}[!h]
	\centering
	\setlength{\tabcolsep}{3.7mm}
	\caption{Timing test of C2P with different algorithms and for
		different grid sizes $n$. The average time per call of
		different C2P schemes (analytic 2d EOS, interpolated 3d EOS,
		NN representation of the EOS, NN representations of the C2P
		are given). In addition, the speedup (acc.) with respect to the
		interpolated EOS is shown. Note that a C2P call transforms all grid~points.
	}
	\begin{tabular}{ccccccccccc}
		\toprule
		\textbf{\boldmath{Grid}} & \textbf{\boldmath{Analyt.}} & \textbf{\boldmath{Interp.}} &
		\multicolumn{2}{c}{\textbf{\boldmath{NNEOSA}}} & \multicolumn{2}{c}{\textbf{\boldmath{NNEOSB}}} &
		\multicolumn{2}{c}{\textbf{\boldmath{NNC2PS}}} & \multicolumn{2}{c}{\textbf{\boldmath{NNC2PL}}} \\
		\textbf{\boldmath{Size}} & \small{\textbf{\boldmath{[$10^{-4}$s]}}} & \small{\textbf{\boldmath{[$10^{-2}$s]}}} 
		& \textbf{\boldmath{[$10^{-3}$s]}} & \textbf{\boldmath{acc.}} & \textbf{\boldmath{[$10^{-3}$s]}} & \textbf{\boldmath{acc.}}
		& \small{\textbf{\boldmath{[$10^{-3}$s]}}} &\textbf{\boldmath{acc.}} & \small{\textbf{\boldmath{[$10^{-3}$s]}}} & \textbf{\boldmath{acc.}} \\
		\midrule
		& & & & & & & & & &\\[-2ex]
		$100$ & $2.45$ & $1.56$
		& $4.94$ & $3.15$
		& $2.69$ & $5.79$
		& $0.98$ & $15.83$
		& $1.23$ & $12.71$ \\
		$200$ & $2.79$ & $2.70$
		& $6.76$ & $3.99$ 
		& $3.65$ & $7.38$
		& $1.12$ & $24.13$
		& $1.90$ & $14.17$ \\
		$400$ & $3.33$ & $4.91$
		& $12.28$ & $4.00$
		& $6.49$ & $7.58$
		& $1.82$ & $27.02$
		& $3.71$ & $13.23$ \\
		$800$ & $4.58$ & $9.39$
		& $22.40$ & $4.19$ 
		& $13.95$ & $6.73$
		& $4.16$ & $22.57$
		& $6.84$ & $13.73$ \\
		$1600$ & $6.64$ & $18.46$
		& $44.05$ & $4.19$
		& $25.67$ & $7.19$
		& $8.46$ & $21.82$ &
		$13.96$ & $13.23$ \\
		$3200$ & $10.99$ & $36.10$
		& $83.86$ & $4.30$ 
		& $49.06$ & $7.36$
		& $16.25$ & $22.21$ 
		& $26.82$ & $13.46$ \\
		\bottomrule 
	\end{tabular}
	\label{tab:C2P:timing}
\end{table}

\subsection{Shock Tube}\label{sec:res:st}

We consider here the 1D shock tube problem discussed in~\cite{Marti.2003} where the initial left/right states are given by $p=(13.33,10^{-6})$ and $\rho=(10,1)$ and the initial velocity is zero. The EOS is given by Equation~\eqref{eq:eos} with $\Gamma=5/3$. The exact solution of the Riemann problem exists and can be computed, e.g.,\ with the method given in references~\cite{Marti.1994,Marti.2003,Mach.2009}. Initial data are evolved with a standard HRSC algorithm using the HLLE flux~\cite{Harten.1983}, linear reconstruction of primitive variables with centered monotonized limiters and a third order Runge--Kutta time integration with Courant--Friedrich--Lewi factor of 0.5. 

\begin{figure}[!h]
	\centering
	\includegraphics[width=.49\linewidth]{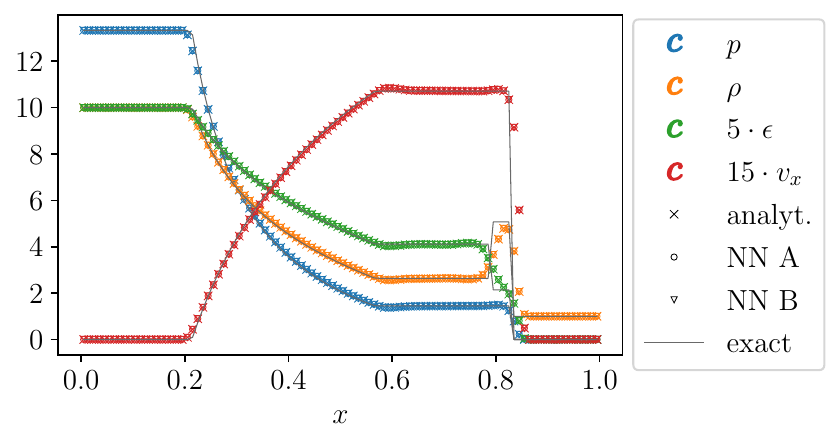}
	\includegraphics[width=.49\linewidth]{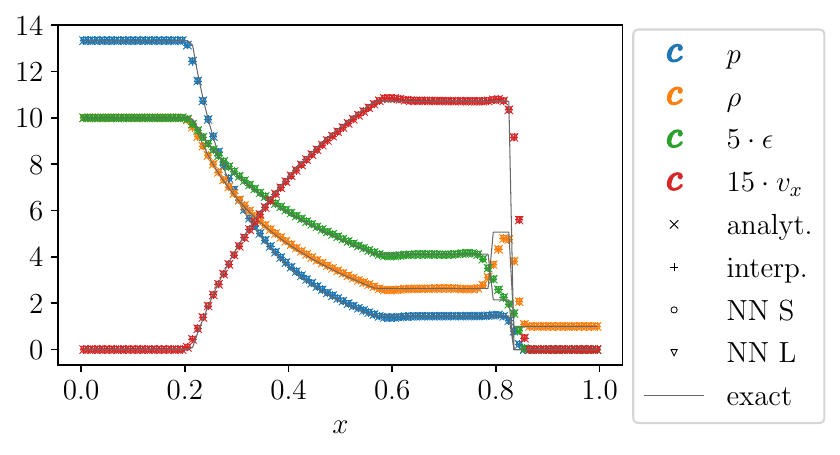}
	\caption{Solution to the shock tube problem at $t\!=\!0.4$ as
		computed by different algorithms. Primitive variables are
		indicated with different colors; the exact solution is
		indicated in solid gray lines. {The y axis represents the value of the primitive variables, rescaled by factors
			shown in the legend for readability.}
		The left panel compares NNEOSA and NNEOSB to the exact
		solution and the solution obtained with the analytical EOS
		representation.
		The right panel compares NNC2PS and NNC2PL to the exact
		solution, the solution obtained with the NR algorithm and the analytical EOS
		representation and the tabular EOS.
		The numerical solutions are computed at a resolution of
		$n\!=\!100$ grid points with step size of
		$\mathrm{d}t\!=\!5\!\cdot\!10^{-3}$.
		{The close agreement between the results
			using NNs and the results obtained with traditional C2P evaluations shows
			that the dominant source of error arises form the HRSC scheme used to
			solve the Euler equations and not the introduction of NNs.}}
	\label{fig:blast1}
\end{figure}

In the left panel of Figure~\ref{fig:blast1} we demonstrate the numerical evolution of the shock tube problem using NNEOSA, NNEOSB and an analytical evaluation of the EOS to recover primitive variables. As a reference, the exact solution of the shock tube problem is also displayed. Here we observe a close agreement between all three approaches and the exact solution.
Similarly, the right panel of Figure~\ref{fig:blast1} compares the performance of NNC2PS and NNC2PL with the analytical evaluation of the equation of state, as well as a comparison with an interpolation of an EOS table. Again the NN approaches provide a close agreement with the exact solution.

\begin{figure}[!h]
	\centering
	\includegraphics[width=.9\linewidth]{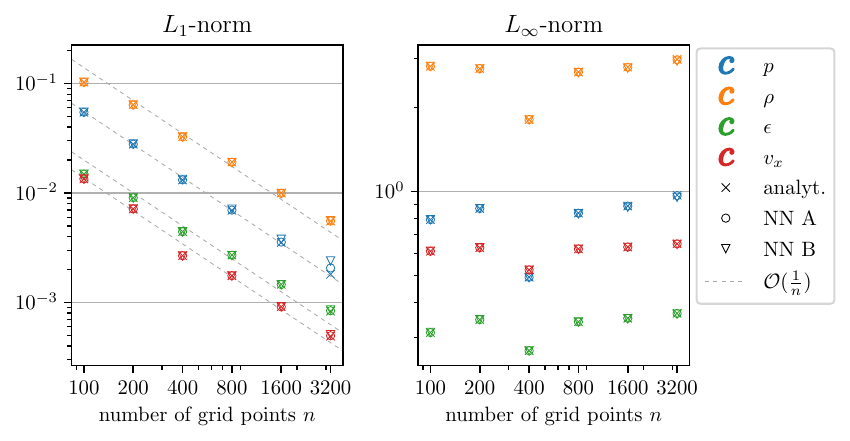}
	\caption{Errors of NNEOS algorithms in the shock tube
		problem. $L_1$ (\textbf{left}) and $L_\infty$ (\textbf{right}) norms of the
		solutions at $t=0.4$ are shown
		for different primitive quantities. First order convergence in
		$L_1$ norm is measured in all 
		quantities and for all algorithms.} 
	\label{fig:Error_eos_blast1}
\end{figure}

\begin{figure}[!h]
	\centering
	\includegraphics[width=.9\linewidth]{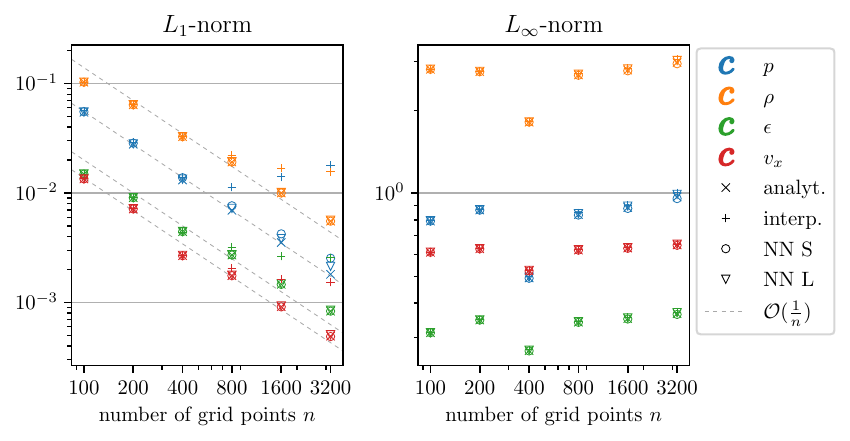}
	\caption{Errors of NNC2P algorithms in the shock tube
		problem. $L_1$ (\textbf{left}) and $L_\infty$ (\textbf{right}) norms of the solutions are shown
		for different primitive quantities. The slope of a first order
		convergence is illustrated by the dashed gray lines in the $L_1$
		plot. First order convergence in $L_1$ norm is measured in all
		quantities and for all algorithms, up to the precision of the tabulated EOS.}
	\label{fig:Error_c2p_blast1}
\end{figure}

The respective errors between these methods are shown in Figures~\ref{fig:Error_eos_blast1} and \ref{fig:Error_c2p_blast1}. Firstly in Figure~\ref{fig:Error_eos_blast1} (left) we demonstrate the $L_1$ norm of the error between the exact solution and the three EOS evaluation approaches used: analytical, NNEOSA and NNEOSB. Consistent with the numerical methods used, and the discontinuous nature of the shock tube profile, we find that the $L_1$ norm converges at first order, as demonstrated by dashed lines. When comparing the two NNs we find that the error in NNEOSA is consistent with that in NNEOSB and in fact for the largest grid sizes is slightly superior in the calculation of the pressure, by ${\sim}17\%$, despite the superior performance detailed in Table~\ref{tab:NNconfigurations}.
This results from the error in the NNs in representing the initial data of the shock tube problem, with NNEOSA representing the initial pressure in the right state with an error of ${\sim}1\cdot10^{-6}$, whereas NNEOSB returns a value deviating several orders of magnitude more (error of ${\sim}5\cdot 10^{-4}$). As demonstrated in Figure~\ref{fig:Error_eos_blast1}, however, any loss of accuracy in replacing the analytic EOS evaluation with either NN is minimal. 

In Figure~\ref{fig:Error_eos_blast1} (right) we observe that, as in the $L_1$-norm, the errors in both NNs and the analytic EOS evaluation are consistent with each other. Here, however, we fail to see the error converging at the expected first order as the resolution increases. This is as a result of the $L_\infty$-norm selecting the point of our domain with maximum error, the location of the shock wave in the shock tube problem (shown at point $x\approx0.83$ in Figure~\ref{fig:blast1}), at which specific point the HRSC algorithm fails to converge.

For the case in which the entire C2P is replaced by an NN, we show the $L_1$-norm of the errors in Figure~\ref{fig:Error_c2p_blast1} (left). Here we again show the expected first order convergence with dashed lines and include the case of the NR
algorithm with the EOS in tabulated form. For lower resolutions (up to 800 grid points) we see first order convergence for all four methods. Beyond this resolution we see the interpolated tabulated EOS fails to converge in all variables due to the resolution of the EOS tables, while both NNC2PS and NNC2PL continue to converge at higher resolutions. For all primitives other than the pressure, we see almost identical performances between the remaining three approaches, whereas for the pressure, we see that, as expected, the analytical evaluation of the EOS is most accurate, followed by NNC2PL and then NNC2PS.

\begin{figure}[!h]
	\centering
	\includegraphics[width=.9\linewidth]{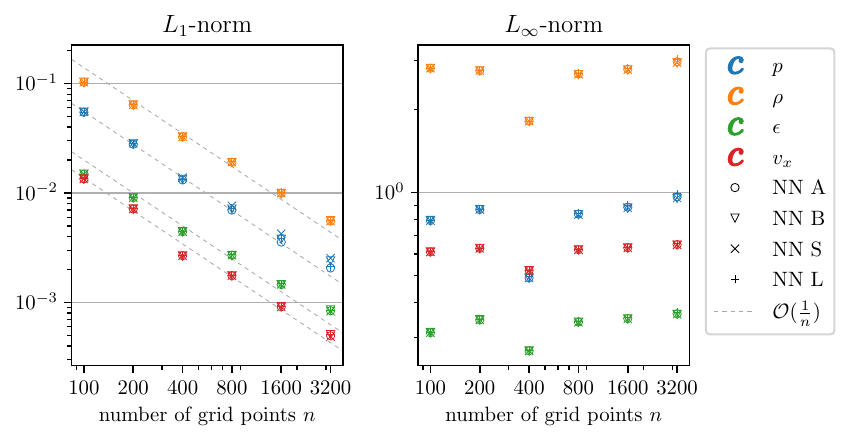}
	\caption{Errors of NNC2P and NNEOS algorithms in the shock tube
		problem. $L_1$ (\textbf{left}) and $L_\infty$ (\textbf{right}) norms of the solutions are shown
		for different primitive quantities. The slope of a first order
		convergence is illustrated by the dashed gray lines in the $L_1$
		plot. First order convergence in $L_1$ norm is measured in all
		quantities and for all algorithms.}
	\label{fig:Error_ALL_blast1}
\end{figure}

In Figure~\ref{fig:Error_c2p_blast1} (right) we demonstrate the error as calculated in the $L_\infty$-norm. As discussed above for the NNEOS approach, due to the failure of convergence of the HRSC algorithm at the shock wave, the error fails to converge in this norm. However, we again see a consistent error between all three schemes demonstrated, and, at the highest resolution, begin to see that NNC2PS slightly outperforms the other approaches when calculating the density and pressure. Finally, in Figure~\ref{fig:Error_ALL_blast1} we compare the errors of all four NN approaches. In  Figure~\ref{fig:Error_ALL_blast1} (left), for the $L_1$-norm we observe that the errors in all methods are of a consistent order of magnitude. For the pressure, we see that at the highest resolution the best performance is found using NNEOSA, followed closely by NNC2PL, with the performance in all other variables and at all other resolutions being indistinguishable in terms of accuracy.

\subsection{Smooth Sine-Wave}\label{sec:res:sw}

In addition to the shock tube problem, we also test this approach on a smooth 
sine wave profile. In contrast to the shock tube problem, the smooth nature of
this test should allow us to investigate the higher order convergence properties
which are absent in the presence of discontinuities formed in the shock tube 
problem. The initial profile of the fluid is given by
\begin{align} \label{eq:sinewave}
	p = 1, \ \ \ \rho &= 1 + 0.2\cdot\sin(2\pi x), \ \ \ v_x = 0.2 \,,
\end{align}
over the interval $x\!\in\![-1,1]$ and periodic boundary conditions are imposed. The same EOS and HRSC algorithm as above is employed for the numerical solution.

\begin{figure}[!h]
	\centering
	\includegraphics[width=.49\linewidth]{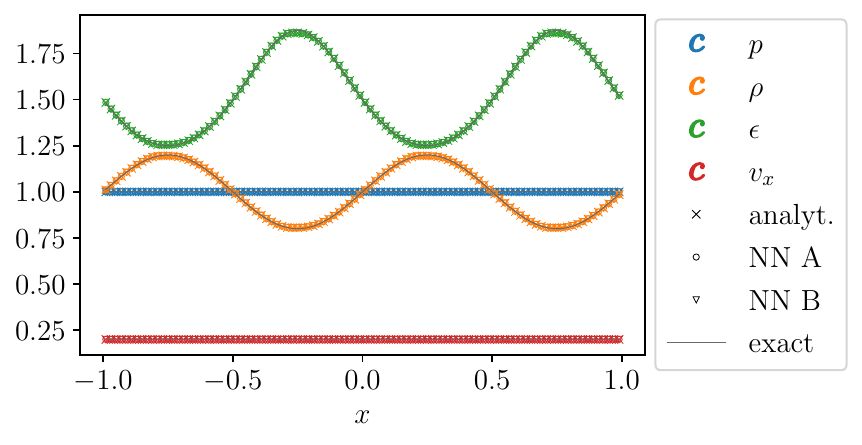}
	\includegraphics[width=.49\linewidth]{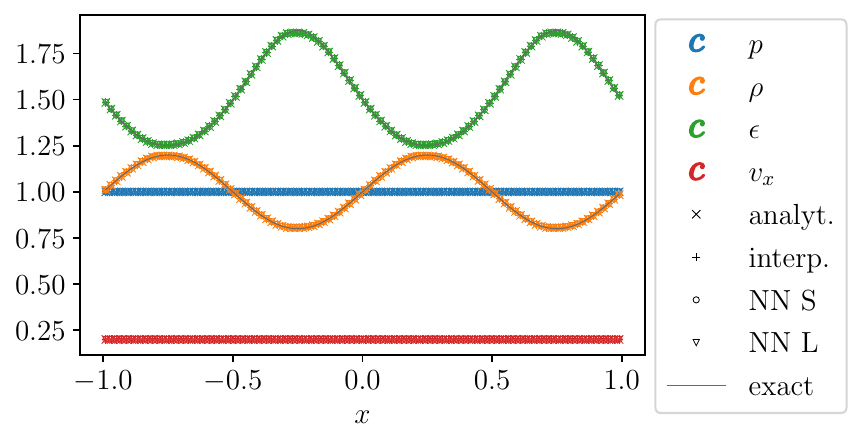}
	\caption{Solution to the sine-wave density profile at $t\!=\!5$
		(i.e., after one period) as computed by different algorithms
		Primitive variables are
		indicated with different colors; the exact solution is
		indicated in solid gray lines.
		The \textbf{left} panel compares NNEOSA and NNEOSB to the exact
		solution and the solution obtained with the analytical EOS
		representation.
		The \textbf{right} panel compares NNC2PS and NNC2PL to the exact
		solution, the solution obtained with the NR algorithm and the analytical EOS
		representation and the tabular EOS.
		The numerical solutions are computed at a resolution of
		$n\!=\!100$ grid points with step size of
		$\mathrm{d}t\!=\!10^{-2}$.}
	\label{fig:sinewave}
\end{figure}

In the left panel of Figure~\ref{fig:sinewave} we demonstrate the agreement between the exact sinewave data after evolving for a single period and the numerically evolved data after the same time calculated with an analytic evaluation of the equation of state, compared with NNEOSA and NNEOSB. Similarly, the right panel of Figure~\ref{fig:sinewave} demonstrates the agreement between the exact sinewave profile and the numerical data obtained through evolution using NNC2PS, NNC2PL, an analytical evaluation of the EOS and through interpolating a tabulated EOS.

\begin{figure}[!h]
	\centering
	\includegraphics[width=.9\linewidth]{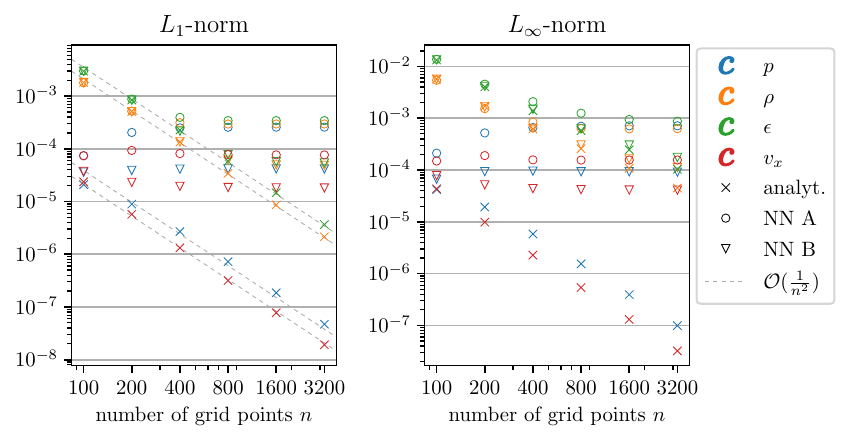}
	\caption{Errors of NNEOS algorithms in the sine-wave density profile 
		problem. $L_1$ (\textbf{left}) and $L_\infty$ (\textbf{right}) norms at $t=5$ of the solutions are shown
		for different primitive quantities. Second order convergence in $L_1$ norm is measured in all
		quantities and for all algorithms until the NN error dominates.}
	\label{fig:Error_eos_sinewave}
\end{figure}

We again demonstrate the error properties of these evolved numerical data in further detail. Firstly the $L_1$-norm of the error for NNEOSA and NNEOSB is shown in Figure~\ref{fig:Error_eos_sinewave} (left). In contrast to the shock tube problem above, the smooth nature of the profile leads us to expect second order convergence, demonstrated by dashed lines. As expected, we see perfect second order convergence when the EOS is evaluated analytically; however, for both NNs we see the error saturate at a floor value (note the absolute values of these errors are considerably smaller than those for the shock tube above). This behavior arises as the error in the EOS representation is independent of the grid resolution, and so, for a small enough grid spacing, will eventually dominate the error arising from the HRSC scheme. The values at which the NN errors dominate, and therefore convergence is lost, are consistent with the accuracies in Table~\ref{tab:NNconfigurations} of $\sim 10^{-4}$ accuracy in pressure. In Figure~\ref{fig:Error_eos_sinewave} (right) we see similar behavior for the $L_\infty$-norm with slightly larger floor values consistent with the slightly larger values reported in Table~\ref{tab:NNconfigurations} for the $L_\infty$-norm. Overall we see that NNEOSB consistently provides smaller errors than NNEOSA. We further note that this floor value in error could be reduced by enlarging the size of the NN, with the trade-off of making the EOS calculation slower and more expensive.
\begin{figure}[!h]
	\centering
	\includegraphics[width=.9\linewidth]{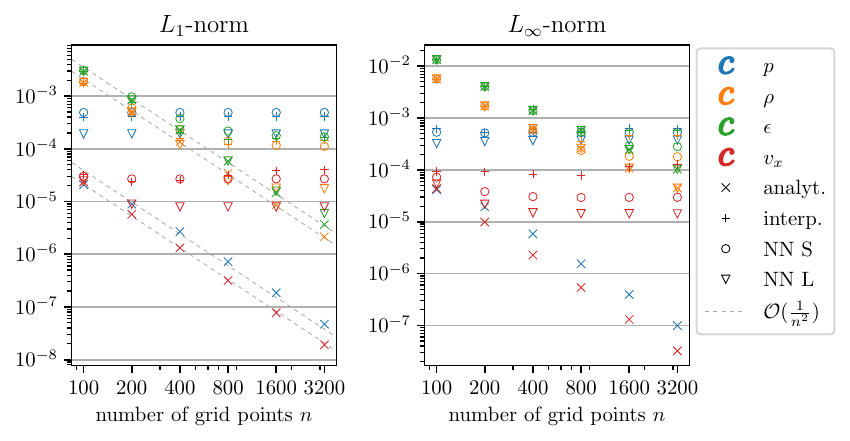}
	\caption{Errors of NNC2P algorithms in the sine-wave density profile  
		problem. $L_1$ (\textbf{left}) and $L_\infty$ (\textbf{right}) norms at $t=5$ of the solutions are shown
		for different primitive quantities. Second order convergence in $L_1$ norm is measured in all
		quantities and for all algorithms until the NN error or precision of the tabulated EOS dominates.} 
	\label{fig:Error_c2p_sinewave}
\end{figure}

For NNC2PS and NNC2PL the $L_1$ norm of the error is shown in Figure~\ref{fig:Error_c2p_sinewave} (left), again with second order convergence denoted by a dashed line. As before, we see that the analytic EOS call converges perfectly, with the NNC2PS and the interpolated tabulated EOS converging until a floor value in the error is reached. In contrast, however, NNC2PL performs better than the other NNs, and, while the error in certain quantities such as the pressure and velocity saturate almost immediately, the error in the internal energy $\epsilon$ still converges and does not appear to have reached its minimum value at the highest resolution used, $n = 3200$. In the $L_1$-norm we find that NNC2PL outperforms NNC2PS in all primitive variables, and, for $n \geq 400$ NNC2PL outperforms the tabulated EOS also. At the highest resolution  we find that NNC2PL outperforms NNC2PS by a factor of ${\sim}2.6$ and the tabulated EOS by a factor of ${\sim}2$ for the error in the pressure and, for the internal energy, which is still converging, this factor increases to ${\sim}29$ compared to NNC2PS and ${\sim}28$ compared to the tabulated EOS.

\begin{figure}[!h]
	\centering
	\includegraphics[width=.9\linewidth]{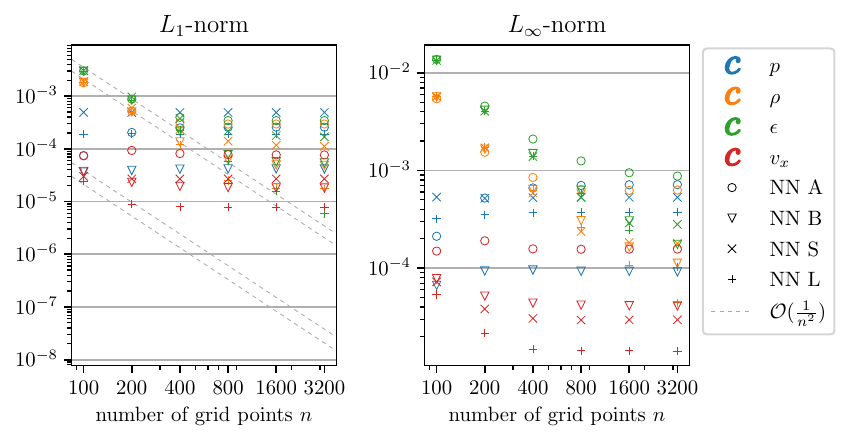}
	\caption{Errors of NNC2P and NNEOS algorithms in the sine-wave density profile  
		problem. $L_1$~(\textbf{left}) and $L_\infty$ (\textbf{right}) norms at $t=5$ of the solutions are shown
		for different primitive quantities.} 
	\label{fig:Error_ALL_sinewave}
\end{figure}

In Figure~\ref{fig:Error_eos_sinewave} we see that even on the coarsest grid used, $n=100$, the error in the NNEOSA representation of the velocity is about a factor three larger than the analytic EOS representation. In contrast, when replacing the entire C2P with an NN, we see in \linebreak Figure~\ref{fig:Error_c2p_sinewave} that at this resolution the velocity error is now comparable to the analytic value. In Figure~\ref{fig:Error_ALL_sinewave} (left) we compare the error between all NN approaches in the $L_1$-norm.  Here we can  see that, once the error has saturated at its floor value for the velocity and density, NNC2PL provides the lowest errors, performing an order of magnitude better than NNEOSA. This is also true for the specific internal energy which has not saturated its error at the highest resolution measured. In contrast, once the error has saturated, NNEOSB gives the lowest error for the pressure, performing an order of magnitude better than NNC2PS and a factor ${\sim}4.5$ better than NNC2PL.

As above we see a similar picture in the $L_\infty$-norm albeit with larger absolute errors.

\vspace{-6pt}


\section{Conclusions}\label{sec:conc}

In this work, we explored artificial neural networks (NN) for the transformation of conservative quantities to primitives (C2P) in relativistic
hydrodynamics. Working in a controlled yet complete framework, we demonstrated that the EOS evaluation or, alternatively, the entire
C2P variables transformation can be substituted by simple and efficient NNs.

Our tests show that an NN representation of the EOS, either outputting the pressure (NNEOSA) or also including pressure derivatives (NNEOSB), can speed up by a factor between 3.1 and 7.4 standard C2P algorithms based on root-finders that are called 3D tabulated EOS. Moreover, the C2P employing NNEOSB runs up to two times faster than the same with NNEOSA, despite being a larger NN. This is due to the cost of the backpropagation needed to compute derivatives for NNEOSA and highlights an advantage of an NN including derivatives as regular output.
The average error reached in the pressure recovery is of order of magnitude $10^{-4}$ for NNEOSA and NNEOSB, an order of magnitude smaller than that obtained with a tabulated EOS of typical size, with NNEOSB also outperforming NNEOSA and the tabulated EOS by an order of magnitude in its maximum error. These accuracies allow the use of the NNs in full hydrodynamical evolutions.
Standard 1D benchmarks in special relativity indicate that the error of the NN EOS representation does not significantly affect solutions of Riemann problems up to very high resolutions of $n=3200$ grid points. This result is expected to hold also in multi-dimensional simulations since in those cases shocks and other elementary waves are resolved with fewer grid points. Note that at such high resolutions table interpolation errors can instead affect the convergence of the solution, as demonstrated by our tests.
Similar considerations hold for smooth solutions. The second order convergence of a smooth flow solution is lost as the errors reach the finite precision of the NN representing the EOS. However, this issue is expected to arise also when using EOS tables and interpolation.
We further note that, since any given physical EOS model (based for example on tabulated data) will only be accurate up to a certain error, an NN based on such a model should only be trained to a matching level of precision. 

Our results indicate that an NN representation of the entire C2P is overall more effective than an NN representation of the EOS solely. In this case, the acceleration of the C2P algorithm is significant when compared to root-finding algorithms with multi-dimensional (3D) table interpolation for the EOS. In the discussed benchmarks, the speedup for the C2P is about a factor ${\sim}13$ for the larger NN (NNC2PL) up to ${\sim}22$ for the smaller NN (NNC2PS), thus depending mainly on the NN size. Our NNs aim to balance speed and accuracy, though further experimentation may be able to further improve this performance. NNC2PL significantly exceeded the accuracy of the solutions with EOS table interpolation in both shock-tube and sinewave benchmarks, with all primitives calculated more accurately up to an improvement by a factor of $\sim$28 still allowing a speed-up of more than an order of magnitude. The benchmark tests also show that, for a shock-tube problem with larger overall errors, the two approaches of NNEOS and NNC2P give similar errors, with NNEOSA improving the accuracy of NNC2PL by ${\sim}4\%$. However, for a sinewave problem with smaller overall errors, NNC2PL gives the most accurate calculation of the velocity, density and internal energy, outperforming NNEOSB by up to a factor of $\sim$9.2, while NNEOSB performs best on the pressure outperforming NNC2PL by a factor of $\sim$4.5. We also remark that the time needed for a C2P using a standard root finder also depends on the problem it is applied on: large and sudden changes, e.g.,\ of shock waves and other discontinuities, cause more iterations of the root-finding algorithms and thus need more time. An NN representation of the C2P does not suffer from this drawback and thus can also improve the robustness of the simulation.

In summary, the results reported in this paper indicate clear advantages in including machine learning representations of the C2P in relativistic hydrodynamics. 
Going beyond these preliminary tests, future work will investigate the applicability of our findings to full multi-dimensional simulations in relativistic astrophysics.


\vspace{6pt} 



\authorcontributions{Conceptualization, T.D., S.B. and D.R.; software, T.D. and S.B.; investigation, T.D.; writing---original draft preparation, T.D., W.C. and S.B.; writing---review and editing, W.C. and D.R.; supervision, W.C. and S.B.; project administration, W.C., S.B. and D.R.; All authors have read and agreed to the published version of the manuscript.}

\funding{S.B. acknowledges support by the EU H2020 under ERC Starting Grant, no.~BinGraSp-714626.
	D.R.~acknowledges support by the U.S. Department of Energy, Office of Science, Division of Nuclear Physics under Award Number(s) DE-SC0021177 and from the National Science Foundation under Grants No. PHY-2011725 and PHY-2116686.
}

\dataavailability{All data are available in the publication. The codes produced for the publication are available on reasonable request.}

\acknowledgments{The authors thank A. Perego and K. Wong for discussions that eventually triggered this project and O.Zelenka for ideas for NNEOS implementations.}

\conflictsofinterest{The authors declare no conflict of interest.}


\abbreviations{Abbreviations}{
	The following abbreviations are used in this manuscript:\\
	
	\noindent 
	\begin{tabular}{@{}ll}
		C2P & conservative-to-primitive\\
		EOS & Equation of State\\
		HRSC & High-resolution shock-capturing\\
		NN & Neural network\\
\end{tabular}}

\appendixtitles{yes} 
\appendixstart
\appendix

\section{Conservative-to-Primitive Transformation}
\label{app:c2p:root}

Given an EOS $\bar{p}(\rho,\epsilon)$, the numerical computation of primitive
quantities can be performed by nonlinear root-finders, e.g.,\ a
Newton--Raphson algorithm. In the context of relativistic astrophysics
a common general procedure is to determine the pressure $p$ by
searching the zero of the function~\cite{Marti.2003}
\be \label{eq:rootfinding:f}
f(p) = \bar{p}(\rho_*(p),\epsilon_*(p)) - p\,,
\ee
where from Equation~\eqref{eq:p2c} one obtains the expressions
\begin{align}
	\allowdisplaybreaks
	\rho_*(p) &= \frac{D}{W_*(p)} \label{eq:cp2:rho}\\
	\epsilon_*(p) &= \frac{\tau + D\left[ 1-W_*(p) \right] + p\left[ 1-W_*^2(p) \right]}{DW_*(p)} \label{eq:cp2:eps}\\
	W_*(p) &= \frac{1}{\sqrt{1 - v_*^2(p)}} \\
	v_*^i(p) &= \frac{S^i}{\tau + D + p}\ . \label{eq:cp2:v}
\end{align}

Derivatives of the pressure $\chi:=\p p/\p \rho$ and $\kappa:=\p
p/\p\epsilon$ are required both in the Newton--Raphson algorithm
and for the calculation of the speed of sound, 
\be
c_s^2 = \frac{1}{h}\left( \chi + \frac{p}{\rho^2}\kappa \right)\,, \label{eq:soundspeed}
\ee
entering the numerical fluxes to integrate Equation~\eqref{eq:hyperbolic}.

In astrophysical applications that employ microphysical EOS like neutron star mergers, the pressure
is interpolated from an EOS table in the form of $\bar{p}(\rho,T,Y_e)$,
where the specific internal energy is implicitly given by the temperature
$T$. Hence, each root{-}finding step for~\eqref{eq:rootfinding:f} includes another root finder for the function
\begin{align}
	g(T) = \epsilon_*(T) - \epsilon
\end{align}
in order to find the temperature~\cite{Wilson.2010}.
It is common to interpolate the three-dimensional EOS tables using a
linear scheme. In order to evaluate the EOS at $(\rho,T,Y_e)$, first
the nearest neighbors to the point are determined, then the value
$\bar{p}(\rho,T,Y_e)$ is determined by linear Lagrangian interpolation, using  an eight-point stencil.
For our benchmarks, we adapted this linear
interpolation algorithm from~\cite{OConnor.2010}.

\section{Examples of the Training of NN Representations} \label{app:training}

\begin{figure}[H]
	\centering
	\includegraphics[width=.49\linewidth]{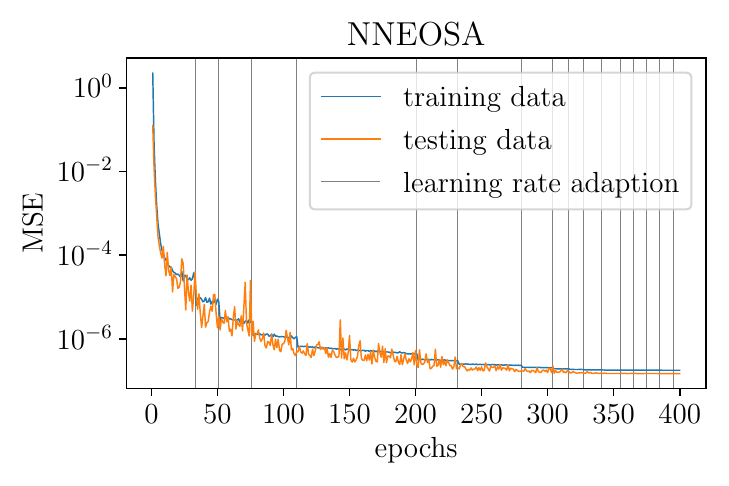}
	\includegraphics[width=.49\linewidth]{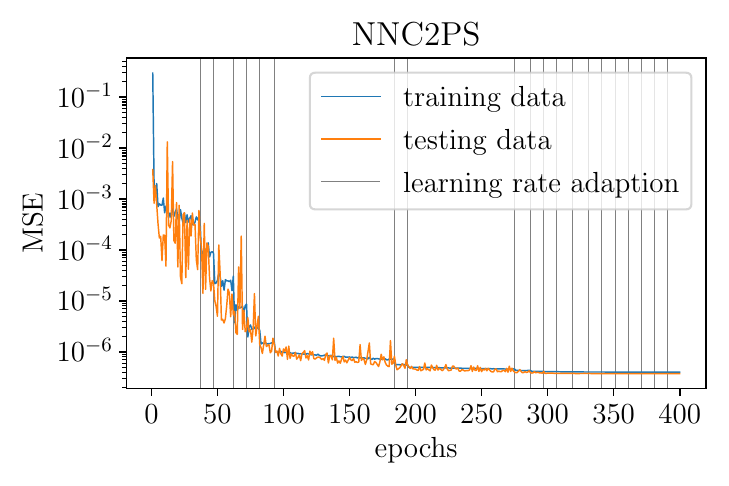}
	\caption{{Training progress of NNs of type NNEOSA (\textbf{left}) and NNC2PS (\textbf{right}). The mean squared errors (MSE) of the outputs to the training data of each epoch are depicted in blue. In addition, the MSE on the testing data after each epoch is shown. The gray vertical lines mark learning rate adaptions.}}
	\label{fig:training}
\end{figure}

Figure \ref{fig:training} visualizes the training progress of NN representations of type NNEOSA (left) and NNC2PS (right) (see Table \ref{tab:NNconfigurations} for hyperparameters). The mean squared error (MSE,~(\ref{eq:MSE})) on the training data is depicted for each epoch as well as the MSE on the testing data computed after each epoch. The initial learning rate of the Adam optimizer~\cite{Kingma.22.12.2014} is set to $1\cdot10^{-4}$ in case of NNEOSA and $6\cdot10^{-4}$ for NNC2PS. The learning rate is later multiplied with a factor of $0.5$ whenever the loss of the training data over the last five epochs does not improve by at least $0.05\%$ with respect to the previous five epochs. Ten epochs have to be completed before the next possible learning rate adaption. The learning rate adaptions are marked by the gray vertical lines in Figure \ref{fig:training}.

The training of NNEOSB and both NNC2PLs is carried out in a similar manner.
\vspace{-6pt}

\reftitle{References}


\end{document}